\documentclass[%
aps, superscriptaddress,
amsmath,amssymb,
preprint,%
]{revtex4-1}

\usepackage{graphicx}
\usepackage{dcolumn}
\usepackage{bm}

\begin{document}


\title{Scanning micro-resonator direct-comb absolute spectroscopy}

\author{Alessio Gambetta}
\affiliation{Dipartimento di Fisica - Politecnico di Milano, Piazza Leonardo da Vinci 32, 20133 Milano, Italy}

\author{Marco Cassinerio}
\affiliation{Istituto di Fotonica e Nanotecnologie - CNR, Piazza Leonardo da Vinci 32, 20133 Milano, Italy}

\author{Davide Gatti}
\affiliation{Istituto di Fotonica e Nanotecnologie - CNR, Piazza Leonardo da Vinci 32, 20133 Milano, Italy}

\author{Paolo Laporta}
\affiliation{Dipartimento di Fisica - Politecnico di Milano, Piazza Leonardo da Vinci 32, 20133 Milano, Italy}
\affiliation{Istituto di Fotonica e Nanotecnologie - CNR, Piazza Leonardo da Vinci 32, 20133 Milano, Italy}

\author{Gianluca Galzerano}
\email{gianluca.galzerano@polimi.it \\Dedicated to Prof. Orazio Svelto, on the occasion of his 80$^{\textrm{th}}$ birthday}
\affiliation{Istituto di Fotonica e Nanotecnologie - CNR, Piazza Leonardo da Vinci 32, 20133 Milano, Italy}
\affiliation{Dipartimento di Fisica - Politecnico di Milano, Piazza Leonardo da Vinci 32, 20133 Milano, Italy}

\date{\today}

\begin{abstract}
Direct optical frequency Comb Spectroscopy (DCS) is proving to be a fundamental tool in many areas of science and technology thanks to its unique performance in terms of ultra-broadband, high-speed detection and frequency accuracy, allowing for high-fidelity mapping of atomic and molecular energy structure. Here we present a novel DCS approach based on a scanning Fabry-P\'erot micro-cavity resonator (SMART) providing a simple, compact and accurate method to resolve the mode structure of an optical frequency comb. The SMART approach, while drastically reducing system complexity, allows for a straightforward absolute calibration of the optical-frequency axis with an ultimate resolution limited by the micro-resonator resonance linewidth and can be used in any spectral region from XUV to THz. An application to high-precision spectroscopy of acetylene at 1.54 $\mu$m is presented, demonstrating frequency resolution as low as 20~MHz with a single-scan optical bandwidth up to 1~THz in 20-ms measurement time and a noise-equivalent-absorption level per comb mode of 2.7$\cdot10^{-9}$~cm$^{-1}$~Hz$^{-1/2}$. Using higher finesse micro-resonators along with an enhancement cavity, this technique has the potential to improve by more than one order of magnitude the noise equivalent absorption in a multiterahertz spectral interval with unchanged frequency resolution. 
\end{abstract}

\maketitle

\noindent Over the past decade high-resolution and broadband spectroscopy has received a major boost from the advent of optical frequency combs (OFCs), highly-coherent light sources constituted by an array of evenly spaced optical narrow lines whose absolute frequencies are known with a fractional accuracy of 10$^{-15}$ or even better \cite{Cundiff2008, Udem2009f}. Direct comb spectroscopy (DCS) exploits high-speed (or parallel) approaches in order to simultaneously detect such a massive set of extremely accurate channels, allowing for ultra-broadband and high speed spectroscopic investigations \cite{Udem2009, Foltynowicz2011, Newbury2011}. Since the first DCS experiment  \cite{Marian2004}, several detection methods have been developed in order to increase the frequency resolution without sacrificing the broadband requirements  \cite{Keilmann2004, Marian2005, Thorpe2006, Diddams2007, Gohle2007, Thorpe2008, Mandon2009, Bernhardt2010, Foltynowicz2011b, Cingoz2012}.  Only two DCS techniques have however demonstrated a frequency resolution down to the comb-tooth level. The first method is the dual-comb or coherent multi-heterodyne spectroscopy, employing two OFC sources with slightly detuned comb-line spacing \cite{Schiller2002, Keilmann2004} that produce an interferogram in the time domain. Spectroscopic informations on the absorption features of the sample are retrieved by Fourier-transform of the interferogram. While providing real-time acquisition capabilities, this technique relies on sophisticated frequency stabilization servo loops or adaptive real time sampling to overcome the frequency noise and relative drifts between the two OFCs \cite{Coddington2008, Roy2012, Ideguchi2014, Rieker2014}. State-of-the-art systems in the near infrared region have shown to resolve the single comb tooth with a frequency resolution of 200~kHz over optical bandwidths as broad as 10~THz  \cite{Ideguchi2014}. Very recently, a compact dual-comb system based on intensity modulated CW laser has demonstrated an even better resolution of 13~kHz over an optical bandwidth of 0.3~THz \cite{Millot2016}. A second  approach is based on the combination of two orthogonal spectral-dispersing elements projecting the resolved modes of a single comb onto a bi-dimensional optical sensor such a CCD. By exploiting a virtually imaged phased array (VIPA) \cite{Shirasaki1996} as a highly dispersing element, several implementations have been proposed able to detect single comb lines with a frequency resolution in the range 0.6 to 1~GHz \cite{Diddams2007, Nugent2012, Berghaus2015, Kowzan2016}. A variant of the approach is based on the substitution of the VIPA with a Fabry-P\'erot (FP) resonator \cite{Gohle2007, Thorpe2008}. The resonator modes are scanned across the frequency comb like a Vernier in frequency space and groups of comb lines containing the spectroscopic information are streaked on a two-dimensional detector array through the use of a scanning mirror synchronized to the cavity sweep cycle. In principle this technique allows for extremely high resolutions, down to the Hz level, however several technical shortcomings mainly ascribable to the finite resolution of the CCD camera limit its spectral resolution to 1.1 GHz \cite{Zhu2014}. The use of a CCD camera also prevents this technique to be adopted in spectral regions where fast, high-resolution imaging devices are not available (XUV and THz), while the unavoidable aliasing due to the Vernier effect does not allow for a straightforward absolute calibration of the optical-frequency axis. Finally, the cavity length needs to be matched to the optical frequency comb-mode spacing in order to achieve the desired Vernier ratio. This makes the design of the detection stage tightly connected to the characteristics of the laser source employed, feature that may be undesirable in remote sensing applications. 

In this letter, we present a different DCS approach based on a scanning FP micro-cavity resonator, SMART DCS (Scanning Micro-cAvity ResonaTor DCS), that is able to detect the spectrum of the OFC with a resolution much better than the comb-mode spacing, limited by the micro-resonator resonance linewidth. The SMART DCS significantly reduces the system complexity and allows direct implementation of this method to any spectral region from THz to XUV.  

\section{SMART DCS}

\begin{figure}
\centerline{\includegraphics[width=8.5cm]{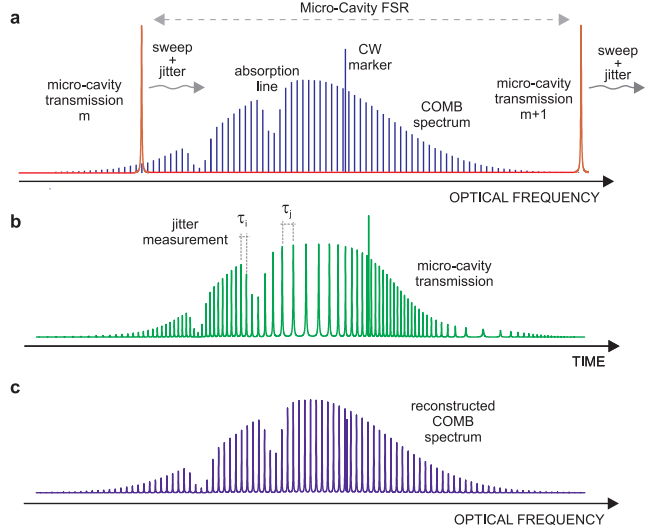}}
\caption{\label{principle}Operating principle of the SMART-DCS method. \textbf{a}, The comb optical spectrum (blu line) is sampled in time by the micro-cavity transmission resonance (red line). The cavity sweep is affected by environmental and electronic noise. \textbf{b}, Micro-cavity transmission over scanning time. The comb line spacing in time ($\tau$) is exploited to evaluate the jitter value along the measurement. \textbf{c}, Optical spectrum of the comb as retrieved after jitter correction and absolute calibration.}
\end{figure}

The SMART DCS approach combines in an efficient way the unique characteristics of the OFC source with the flexibility of a scanning FP interferometer. Figure \ref{principle} sketches the operating scheme. An OFC, whose output spectrum is constituted by evenly spaced optical lines at frequencies $\nu_{comb}=f_0+nf_r$, where $f_0$ is the comb offset frequency, $f_r$ is the comb line spacing, and $n$ is the order of the comb line, is coupled to a FP micro-resonator with a free-spectral-range $FSR=c/(2d)\gg f_r$, where $d$ is the mirrors separation and $c$ is the speed of light. The FP micro-resonator finesse, $\mathcal{F}=\pi\sqrt{R}/(1-R)$ where $R$ is the reflectivity of the cavity mirrors, is such that $\frac{FSR}{\mathcal{F}}=\Delta\nu_{FP}\ll f_r$ corresponding to the case where only one comb mode is resonant with the cavity at a time. The FP transmission versus scanning time corresponds therefore to the classical optical spectrum analyzer configuration, which represents the spectral convolution between the FP Airy function and the comb spectrum. In order to avoid spectral aliasing between the adjacent transmission orders of the FP, the optical bandwidth of the OFC must be narrower than the cavity FSR, condition that may not be fulfilled for many typical OFCs even for a very short spacing of the cavity mirrors. For this reason an optical-tunable-filter (OTF), either before or after the FP micro-resonator, has to be used in order to select the appropriate spectral portion of the OFC to be detected. To further extend the optical bandwidth while maintaining the same measurement time and frequency resolution, the OTF may be replaced by a grating and a photodiode array (each pixel of the array analyzing a single micro-cavity $FSR$) placed at the end of the detection chain. The latter configuration would allow for a simultaneous detection of the intensity transmitted at different orders of the micro-cavity, greatly increasing the single-scan optical bandwidth. 

As sketched in Fig.~\ref{principle}, environmental and electrical noise translates into a temporal jitter of the micro-cavity transmission modes. This effect is amplified by the very short cavity length, being the cavity detuning $\Delta\nu_{FP}=\frac{\Delta d}{d}\nu_{FP}$, and prevents the frequency axis from being linearly mapped onto the acquisition time scale. A correction algorithm is therefore needed to compensate for cavity jitter and reconstruct the optical frequency axis. This is done in post-processing by taking advantage of the clearly resolved comb modes which act as a reference ruler for frequency remapping. For the purpose of absolute frequency calibration of the optical axis a free-running CW laser, with a frequency stability better than the comb spacing, is combined with the OFC. This CW source acts as a frequency marker to measure in real time the comb-mode order, given by the simple relation $N=(\nu_{CW}-f_0)/f_r$.  Exploiting the absolute frequency scale of the OFC, the position of all the measured comb teeth is known with the accuracy of the comb source even in the presence of micro-resonator dispersion, CW laser frequency fluctuations, and cavity length jitter during the scanning time. Therefore, effective coherent averaging of the recorded spectra can be performed to boost the signal-to-noise ratio (SNR) of the detection system and to improve the sensitivity of the technique.

\begin{figure}
\centerline{\includegraphics[width=8.5cm]{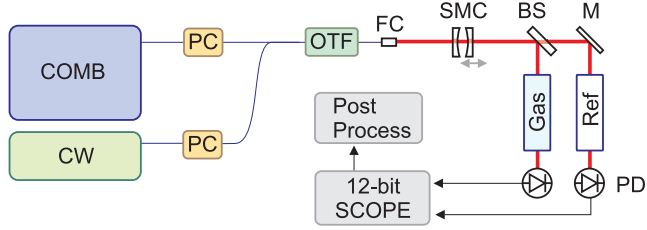}}
\caption{\label{setup}Experimental setup of the SMART DCS method using an Er-fiber OFC at 1.55 $\mu$m. CW: extended cavity laser diode. COMB: self-referenced optical frequency comb. PC: polarization control. OTF: optical tunable filter. FC: free space fiber coupler. SMC: scanning micro-cavity. BS: beam splitter. M: Mirror. Gas: C$_2$H$_2$ filled cell. Ref: empty reference cell. PD: fast InGaAs photo-detector. Cavity transmission vs time is acquired by a 12-bit digital oscilloscope and sent to a Matlab-equipped personal-computer for post processing.}
\end{figure}

\section{SMART DCS of C$_2$H$_2$ at 1.55~$\mu$m}
Feasibility of the SMART DCS has been demonstrated by high-precision spectroscopy of acetylene at around 1.55 $\mu$m. The comb source adopted is a self-referenced Er:fiber laser frequency comb providing a 0.5~W output power, with a 250~MHz pulse repetition frequency, stabilized against a GPS-disciplined Rb Radio-Frequency standard. The measurement setup is sketched in Fig.~\ref{setup}. The comb is combined with the output of a tunable CW laser with $\sim$60~MHz frequency accuracy, filtered by the OTF (optical bandwidth variable in the range 0.5 to 20~nm), coupled to a high finesse ($\sim$50,000) FP micro-cavity and, after passing through a Herriot-type multipass absorption cell (33-m interaction length) filled with C$_2$H$_2$, is shined onto a 2~MHz  bandwidth InGaAs photo-detector. A reference arm constituted by an empty cell and an identical detector is used for calibration, normalization and background removal. A spacing of  $\sim$150~$\mu$m between the cavity mirrors ($T=0.005$\% at 1540 nm) ensures a free-spectral-range limited bandwidth of the order of 1~THz or 8~nm at a central wavelength of 1.55~$\mu$m. The resonance frequency sweep is performed by continuously scanning the cavity-mirrors distance through a periodic voltage ramp applied to a piezo-electric transducer. As stated before, the micro-cavity scan is free-running and completely independent of comb repetition frequency $f_r$. 

\begin{figure}
\centerline{\includegraphics[width=8.5cm]{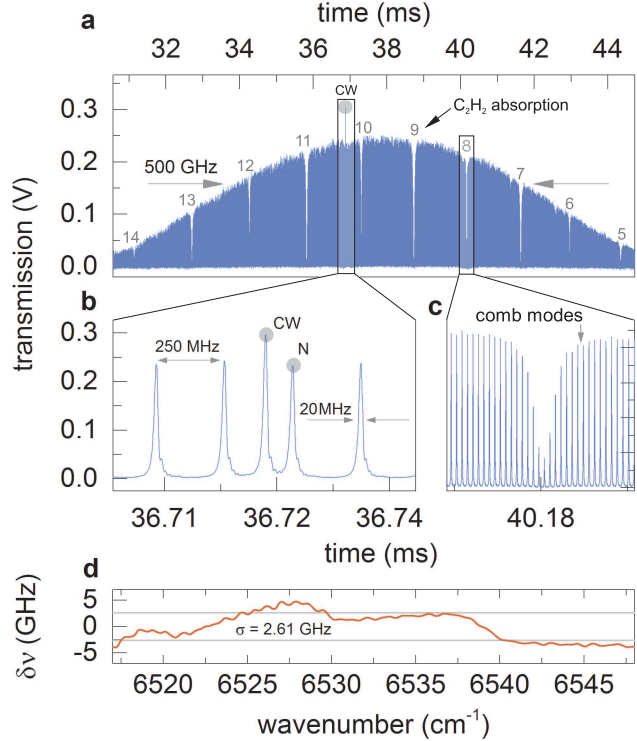}}
\caption{\label{single}\textbf{a}, Micro-resonator transmission versus scanning time. Both OFC and CW laser spectra are detected in the time-trace. A 10-cm long cell filled with C$_2$H$_2$ at 50~mbar is placed along the beam path. Several lines (numbered from 5 to 14) of the C$_2$H$_2$ P-branch absorption band are clearly visible as a modulation of the OFC spectrum. Middle panels: \textbf{b}, temporal zoom around the CW laser showing the 20-MHz frequency resolution (FWHM) and the comb line spacing; \textbf{c}, a detail of the resolved comb modes modulated by the P(8) rotovibrational absorption of the C$_2$H$_2$; \textbf{d}, Micro-resonator jitter measured as the frequency deviation of the comb-peak position from a perfect linear dependence vs optical frequency.}
\end{figure}

\begin{figure}
\centerline{\includegraphics[width=8.5cm]{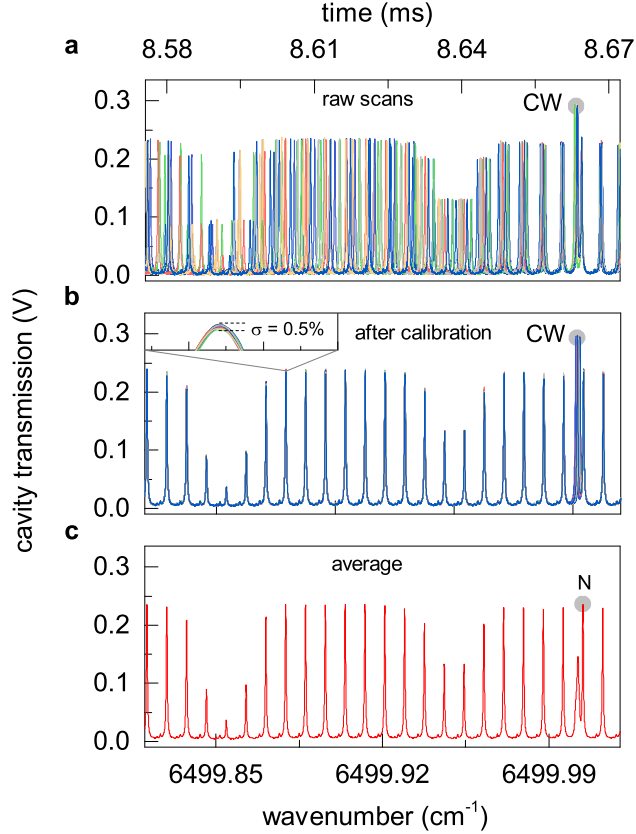}}
\caption{\label{multiple}\textbf{a}, 20 consecutive time-trace acquisitions as recorded by the oscilloscope. Each isolated peak corresponds to a single comb mode. The CW laser line can be distinguished as the highest peak. Two C$_2$H$_2$  absorption features are clearly visible as a modulation of the comb spectrum. \textbf{b}, Measurements after jitter-compensation and absolute frequency calibration. Inset: RMS intensity noise of a comb peak. \textbf{c}, Average of the 20 calibrated measurements.}
\end{figure}

Figure~\ref{single} shows a sample spectrum as acquired through a 12~bit oscilloscope by scanning the cavity length with a span $\Delta d$  of $\sim$0.8~$\mu$m in 14~ms, resulting in a scanning speed of 66~GHz/ms. A final resolution of $\sim$20 MHz can be estimated for the adopted configuration as a result of mirror distance and cavity finesse. Indeed, the individual comb modes as well as the CW laser line are clearly resolved, as shown in Fig~\ref{single}~(b). Absorption features of the C$_2$H$_2$ gas are clearly visible as a modulation of the comb mode intensity (see Fig~\ref{single}~(c)).  A measure of cavity jitter and dispersion, as retrieved by the post processing compensation and reconstruction algorithm, is reported in Fig.~\ref{single}~(d), showing the difference between the effective center frequency of each comb tooth and its value, as retrieved from the oscilloscope time scale without any correction. It is worth noting that without any environmental isolation of the microcavity, the measured frequency jitter shows a root-mean-square (r.m.s.) value of 2.6~GHz in an optical frequency scan of $\sim$930~GHz, corresponding to a r.m.s. cavity length jitter of 2.2~nm. 

Figure~\ref{multiple}~(a) shows a spectral detail around the CW reference frequency of 20 raw consecutive acquisitions recorded with fixed values of comb offset and repetition frequencies ($f_0=30$~MHz and $f_r=250$~MHz). The time base of each measurement has been translated in order to match the position of the $N^{th}$-order comb tooth, next to the CW laser peak. Mainly due to the microcavity jitter, the recorded spectra are not overlapped in time and effective coherent averaging can not be performed. Figure~\ref{multiple}~(b) reports the same 20 measurements after remapping and absolute frequency calibration. A perfect time overlapping of these spectra is obtained with a SNR of the single measurement limited only by the RIN of the comb at a level of about 0.5\%, in a measurement bandwidth from 50~Hz to 2~MHz.  The average of the 20 spectra is plotted in Fig.~\ref{multiple}~(c), showing also the comb mode $N$ used for absolute calibration. In the averaging the CW laser line is smeared out and its peak value is decreased due the frequency noise along each sweep, whereas the comb spectrum shows an unchanged frequency resolution of 20~MHz. The SNR of the averaged spectra increases as the square root of the samples, as ascribed to a white noise contribution in the comb RIN that reduces to a level of $\sim 10^{-3}$ after 20 averages and to $\sim 4\cdot10^{-4}$ after 100 averages. The noise-equivalent absorption coefficient at 1~s time-averaging per comb line, defined as $(L_{opt}\cdot SNR)^{-1}(T/M )^{1/2}$, is 2.7~$\cdot10^{-9}$~cm$^{-1}$~Hz$^{-1/2}$, where $L_{opt}$ (33 m) is the multipass-cell optical path, $SNR$ (250) is the signal-to-noise ratio, $T$ (20~ms) is the measurement time, and $M$ (4000) is the number of comb lines. The noise-equivalent absorption can be further improved by increasing the optical path using an enhancement cavity instead of a multipass-cell (for example 10$^{-10}$~cm$^{-1}$~Hz$^{-1/2}$ could be readily obtained with a 1-km interaction length by an enhancement cavity finesse of $\sim$3,000).

\begin{figure}
\centerline{\includegraphics[width=8.5cm]{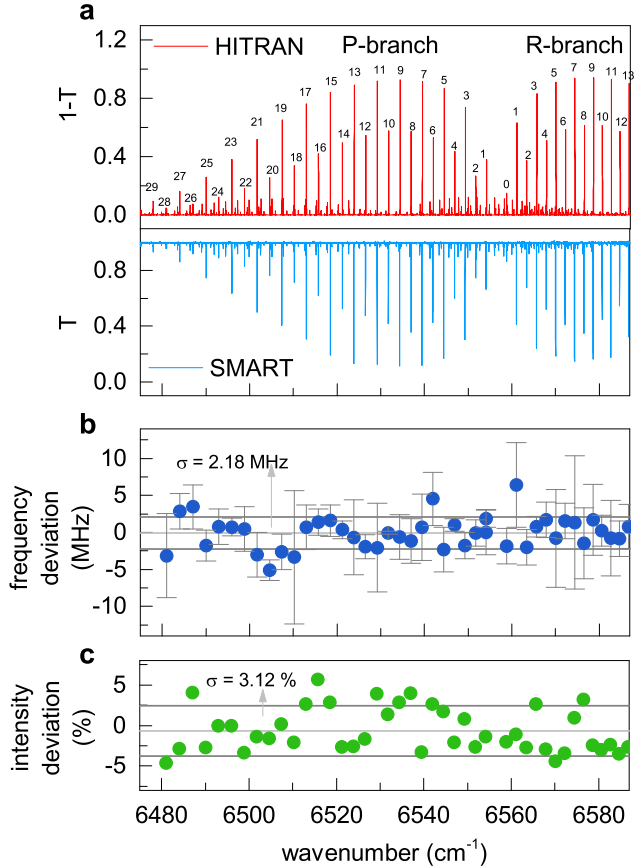}}
\caption{\label{hitran}\textbf{a}, Bottom panel: C$_2$H$_2$ spectrum at room temperature, $p\simeq0.045$~mbar, optical path $L_{opt}$=33~m, obtained via juxtaposition of ten spectrally adjacent SMART measurements. Top panel: HITRAN simulation. \textbf{b}, fractional deviation of the measured peak absorption with respect to the HITRAN simulation showing a rms of 3\%; \textbf{c}, frequency deviation of the retrieved line center frequency with respect to the HITRAN value showing a rms of 2~MHz.}
\end{figure}

\begin{figure}
\centerline{\includegraphics[width=8.5cm]{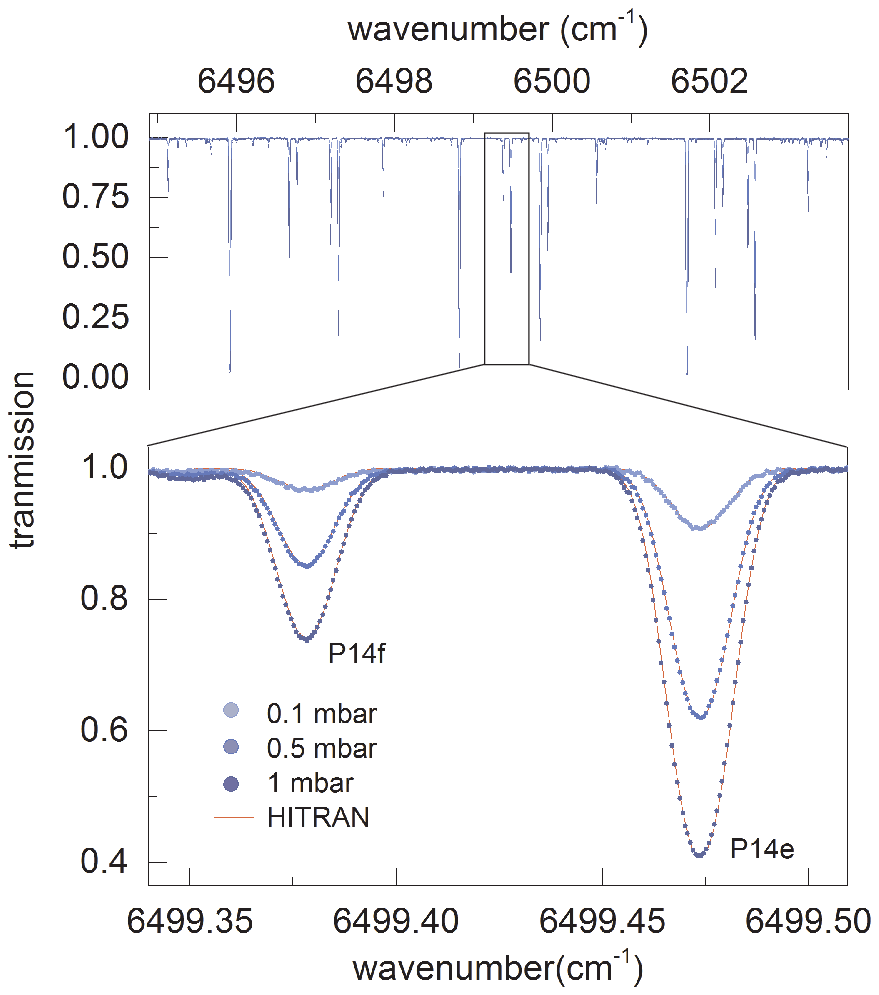}}
\caption{\label{interleaving}High resolution C$_2$H$_2$ SMART-DCS spectrum (room temperature, $L_{opt}$~=~33~m) at three different pressures. Each spectrum is the result of twelve interleaved measurements, with frequency comb spectral detuning of 19.5 MHz between each measurement. \textbf{a}, Full measurement spanning $\sim$9~cm$^{-1}$ (270~GHz); \textbf{b}, particular around the P14f and P14e absorption lines over $\sim$~0.2~cm$^{-1}$ span (dots).  A simulation based on HITRAN databases is also shown (red line). }
\end{figure}

In order to overcome the 8-nm bandwidth limit of the single sweep, consecutive measurements have been performed with the same 1-THz micro-cavity FSR, tuning the center wavelength of the OTF to scan adjacent portions of the spectrum and to extend the overall optical bandwidth investigated. We choose a narrower $\sim$2~nm filter bandwidth for each single-scan in order to keep an almost constant signal-to-noise ratio along the whole span. A C$_2$H$_2$ spectrum at a pressure of $\sim4.5\cdot10^ {-2}$~mbar, obtained by combining 15 measurements performed in 2-nm broad adjacent spectral portions, is shown in Fig.~\ref{hitran}~(a) and compared to the spectrum simulated by HITRAN database \cite{Rothman2013}. Despite the low optical-frequency sampling, due to the 250~MHz comb spacing, very close to the $\sim$450~MHz average linewidth of the absorption feature to be detected, the high SNR and the absolute frequency calibration of the SMART method allow for high-precision measurements of both line intensities and center frequencies. Figures~\ref{hitran}~(b) and (c) show, respectively, the deviation between the measured and computed line center frequency and the fractional deviation between the measured and simulated absorption peaks, showing rms values of 2~MHz and 3\%, respectively, mainly limited by statistical contributions. Features as close as $\sim$600 MHz such as the P1e and R10f lines (respectively at 6554.111497 and 6554.131 cm$^{-1}$) have been detected easily with very good accuracy, with regards to both centerline position (respectively $\sim$0.1~MHz and $\sim$2.1~MHz) and transmission intensity ($\sim$1\% and $\sim$3\% ). 

To overcome the intrinsic resolution limit set by the comb-teeth spacing, sequential interleaved measurements have also been performed by scanning the comb repetition rate. Figure~\ref{interleaving}~(a) shows three C$_2$H$_2$ spectra recorded, respectively, at 0.1~mbar, 0.5~mbar, and 1~mbar. By choosing a frequency detuning of 19.5 MHz, corresponding to a $\Delta f_{rep}$ = 25 Hz, the whole 250 MHz spacing between adjacent comb modes has been covered by twelve consecutive measurements. As shown in Fig.~\ref{interleaving}~(b), the HITRAN simulation agrees very well with the experimental points. The high-precision and resolution performance of the SMART method can be further appreciated by the capability to measure the low pressure-shift coefficients of the investigated C$_2$H$_2$ lines: as a demonstration, an identical coefficient of -0.75(5)~MHz/mbar has been measured for the first time for both P14f and P14e lines of Fig.~\ref{interleaving}~(b), for pressure values up to 50 mbar. Control measurements have been performed on the P22 line, providing a pressure-shift coefficient of -0.388(18)~MHz/mbar, in very good agreement with previous results \cite{Swann2000}.

\section{Conclusions}
We demonstrate a compact and versatile DCS method with a resolution of $\sim$20~MHz and a single-scan bandwith of $\sim$1~THz, capable of absolute frequency measurements with a noise-equivalent-absorption level per comb mode of 2.7$\cdot10^{-9}$~cm$^{-1}$~Hz$^{-1/2}$ in the near-infrared. This method has unique characteristics and performance that allow one to measure molecular absorption profiles with high precision. The detection scheme lends itself to be extended to other spectral regions ranging from XUV to THz and can be adopted in a large variety of experiments from remote sensing to precision spectroscopy. Resolution and bandwidth are further scalable by increasing the resonator finesse, tuning the mirror distance and through parallel detection. In addition, the noise equivalent absorption can be further improved by more than one order of magnitude replacing the multipass absorption cell by an enhancement cavity. The compactness of the spectrometer makes it easily integrable in miniaturized systems, like lab-on-chip devices, especially in combination with micro-ring-based OFC comb sources.

\nocite{*}
%

\section*{Acknowledgements}
The authors acknowledge financial support from the Italian Ministry of University and Research ELI-attosecond infrastructure.

\section*{Author Contributions}
\noindent G.G. and A.G. conceived the concept behind the paper, A.G., M.C., and D.G. conducted the experiments, A.G., P.L. and G.G. wrote the paper and analyzed the results, P.L. and G.G. supervised the work. All authors reviewed the manuscript.

\section*{Additional Information}
Competing financial interests: The authors declare no competing financial interests.

\end{document}